\documentclass[%
 reprint,
 amsmath,amssymb,
 aps,
]{revtex4-2}
\usepackage{tikz}
\usetikzlibrary{shapes.geometric, arrows, positioning}
\usepackage{graphicx}
\usepackage{dcolumn}
\usepackage{bm}
\usepackage{caption}
\usepackage{subcaption}
\usepackage{float}
\usepackage{verbatim} 

\begin{document}

\tikzstyle{startstop} = [rectangle, rounded corners, minimum width=3cm, minimum height=0.5cm,text centered, draw=black, fill=red!30]
\tikzstyle{io} = [trapezium, trapezium left angle=70, trapezium right angle=110, minimum width=2cm, minimum height=0.5cm, text centered, draw=black, fill=blue!30]
\tikzstyle{process} = [rectangle, minimum width=2cm, minimum height=0.5cm, text centered, draw=black, fill=orange!30]
\tikzstyle{decision} = [diamond, minimum width=3cm, minimum height=1cm,inner sep=0pt, text centered, draw=black, fill=green!30]
\tikzstyle{arrow} = [thick,->,>=stealth]
\preprint{APS/123-QED}

\title{Radio-frequency cavity field measurements through free falling bead}
\thanks{A footnote to the article title}%

\author{Xiaonan~Du}
\author{Lars~Groening}%

\affiliation{GSI Helmholtzzentrum für Schwerionenforschung GmbH, Darmstadt, Germany}%


\date{\today}

\begin{abstract}
This paper introduces a novel bead-falling measurement method for the precise and efficient mapping of electromagnetic fields within radio-frequency (RF) cavities, which are crucial components in the design of accelerators. The traditional bead-pull method, while effective, involves mechanical complexities and is prone to errors from wire perturbations. The innovative method reported here leverages the simplicity and accuracy of free-falling beads to mitigate these issues. This technique eliminates the need for a wire-pulley system, thereby simplifying the experimental setup and reducing potential mechanical errors. We detail the development and operational principles of this new method, including the design of a compact, portable measurement device that integrates a bead/droplet release system and a bead detection system linked to a Vector Network Analyzer (VNA). The device has been tested with a three-gap buncher cavity and a scaled Alvarez-type cavity, demonstrating its ability to perform rapid, reliable field measurements under various challenging conditions, including low signal-to-noise ratios and environmental vibrations. The results confirm the method's superiority in precision and operational efficiency, potentially setting a new standard for RF cavity diagnostics and tuning.
\end{abstract}

\maketitle

\section{Introduction}
In the evolving landscape of accelerator physics and the design of resonant cavities, precise measurement and tuning of electromagnetic fields of rf-cavities are paramount. The bead-pull method, a technique steeped in the principles of Slater's small signal perturbation theory~\cite{Maier1952FieldSM,Waldron1960PerturbationTheory}, emerges as a cornerstone in this domain. By introducing a small bead into a resonant cavity, this method allows for the quantification of shifts in the cavity's resonant frequency, attributing these shifts to the perturbative presence of the bead. Such frequency variations are indicative of the electric and magnetic field strengths at the bead's location, offering insights into the field distribution within the cavity.

In a typical bead-pull setup, X-Y linear motion systems are placed on each side for aligning the axis of the wire (fishing line) with the cavity axis. A motorized linear actuator with a specified stroke is used to move the bead along the measurement path inside the cavity by pulling the wire~\cite{Hahn2011HOMIdentification}. In some instances, an optical encoder provides feedback for the bead's location, which is calibrated to ensure precise positioning and movement.

This method not only underpins the characterization of longitudinal and transverse electromagnetic fields within cavities but also facilitates the iterative tuning required to achieve desired field distributions. Its application spans the assessment of cavities crafted from various materials, employing objects like spheres and needles to invoke perturbations. The advent of automated test stands, notably those developed at prestigious institutions like CERN, marks a significant advancement in the method's implementation. These test stands, equipped with motors to maneuver the bead across three axes, enable comprehensive field mapping, ensuring the cavity's optimization for its intended application.

The bead-pull technique's relevance is underscored by its role in the manufacturing and tuning of high-frequency resonant cavities, a critical component of the burgeoning accelerator industry. With the demand for these cavities on the rise, the method's efficacy in facilitating the fine-tuning of electromagnetic field distributions becomes increasingly important. 
The common measurement procedure for Radio-Frequency Quadrupoles (RFQs) consists of measuring each quadrant and comparing field
flatness along the length of the structure. Obtaining the desired field distribution is an iterative process. bead pull measurements were a continuous part of the process, critical for achieving the RFQ's precise electromagnetic field requirements~\cite{PhysRevAccelBeams.20.080102,Berrutti2014PXIERFQBeadPull}.

However, the bead-pull method is not without its limitations. The wire and pulley system typically requires customization for the cavity to be measured, and its final dimensions depend on the cavity under test. The need for mechanical adjustments, dealing with potential issues like bead vibration or wire over-latch, and ensuring stable pulling can make the process more time-consuming and labor-intensive. Additionally, the wire can cause perturbations, particularly concerning the potential interference of the wire system with certain (dipole) modes of the cavity, thereby rendering the proportion of higher-order field components in the measured field distribution inaccurate~\cite{Goudket2008Comparison}.

In pursuit of an innovative approach, we have been developing what we term the bead-falling-measurement method. This technique leverages the free fall of a bead to induce perturbations, thereby eliminating the need for a pulley-wire system and its associated complications. This method presents a promising alternative to the conventional bead-pull measurement, offering an avenue for further refinement and efficiency in cavity diagnostics and tuning.
As the field of accelerator technology continues to advance, the evolution of measurement techniques like the bead-pull and bead-falling methods will play a critical role in shaping the future of resonant cavity measurement and tuning.

The next section illustrates the method's working principle and outlines its advantages towards the conventional technique. It is followed by description of the applied from-the-shelf hardware. The fourth section is dedicated to the applied software. Measurements done with a buncher cavity and a model of an accelerating cavity are presented afterwards. Final conclusions evaluate the method's capabilities and its further potential. Two appendices are on issues related to free-fall including air resistance and to the reduction of noise.

\section{conceptual design and advantages}
\begin{figure}[H]
    \centering
    \includegraphics[width=8 cm]{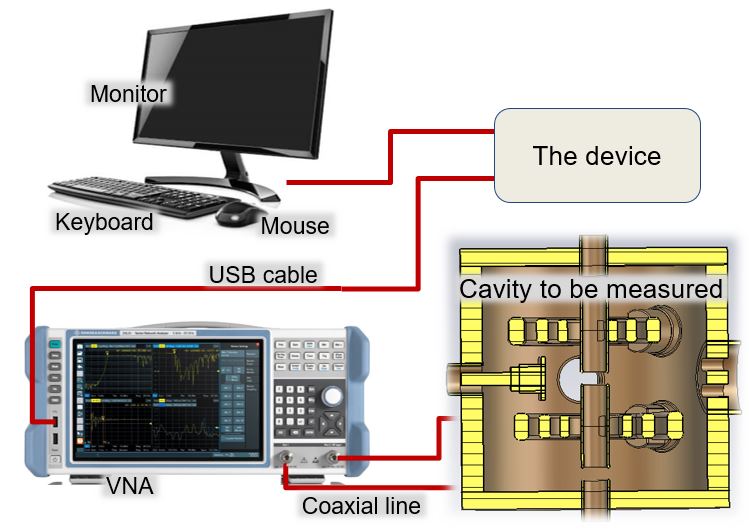}
    \caption{Scheme of bead-fall RF-measurements. The cavity is connected to two of the VNA ports. The device includes the bead/droplet release system and the bead detector. It is positioned such that the falling bead is aligned to the beam axis of the cavity. Beads are released from the top and fall freely through the bead-detector and subsequently through the cavity.}
    \label{fig:conceptual_design}
\end{figure}

Similar to bead-pull measurements, our measurement also requires the use of a bead to create perturbations and employs a Vector Network Analyzer (VNA) to capture Radio-Frequeny (RF)-measurement data. Since the bead is in free-fall through the cavity, the acceleration causes the measured perturbation-time curve to not linearly translate to a perturbation-position curve. Therefore, a setup is needed to release the bead steadily, and then through measurement and calculation, correlate the time-position relationship accurately. Figure~\ref{fig:conceptual_design} shows the conceptual design of this measurement method, and subsequent sections will elaborate on the implementation details.

The new method overcomes several technical challenges inherent in traditional bead-pull setups and brings numerous benefits listed below:
\begin{itemize}
    \item \textbf{well-defined perturbation position}: The bead can be very small (e.g., 0.5 to 1~mm in diameter), ensuring the perturbation position to be precisely defined. This allows the bead to pass through small gaps, like those between RFQ vane tips~\cite{Berrutti2014PXIERFQBeadPull} or X-band cavities~\cite{PhysRevAccelBeams.21.061001} without significant field variation over its size.
    \item \textbf{no perturbation from wire}: Perturbation from the nylon ($\epsilon_r=4$) wire is inevitable in the bead-pull method, though it is not an issue using the new method.
    \item \textbf{fast measurements}: With the free-falling duration of less than a second, measurements can be conducted rapidly. This enables real-time field monitoring at about~2~Hz, allowing tuners to be adjusted based on continuously updated field distributions, thus making the cavity tuning process more efficient.
    \item \textbf{material versatility}: use of different materials for the perturbation, such as liquid drops, facilitates fast iterative bead release, adding flexibility to the measurement process.
    \item \textbf{well-defined path}: The bead's path is determined by the launching device's position and velocity, which can be easily controlled, eliminating the need for a wire pulling device and, thus, avoiding wire and bead vibration as well as perturbation from materials like nylon.
    \item \textbf{simplified setup}: The absence of a wire and pulley system not only simplifies the experimental setup but also significantly saves time during preparation.
    \item \textbf{accessible measurements in complex locations}: The possibility of the bead's path being a parabola, may allow for measurements in cavity locations not accessible by traditional bead-pull measurements.
\end{itemize}

Integrating these advantages, the new method streamlines the experimental setup, enhances measurement speed and accuracy, offers flexibility in measurement, reduces mechanical interference, and enables real-time monitoring. These improvements make the cavity tuning process significantly more efficient and adaptable to varying experimental conditions.

One of the most promising applications of this device is the tuning of RF-cavities. For multi-gap cavities, such as RFQs and Drift Tube Linacs (DTLs), a complex field distribution tuning process is always necessary to address deviations in the actual electric field distribution caused by mechanical machining errors from the design value (typically equal electric field peaks in all cells). This tuning process usually requires multiple field distribution measurements using the bead pull measurement technique, thereby iteratively adjusting the field distribution closer to the design value.

First, the device introduced in this paper can significantly simplify the experimental preparation phase, as it integrates all necessary functionalities into a small box, requiring only a single alignment on top of the cavity. Second, it enables each field distribution measurement to be completed in a very short time (less than~0.5~s), and continuous measurements can be conducted at these intervals to achieve real-time monitoring of the field distribution. Any changes in the field distribution caused by moving the tuner can be immediately presented to the experimenters. This can greatly accelerate the entire field tuning process.

\section{Hardware}
\begin{figure}[H]
    \centering
    \includegraphics[width=8.5cm]{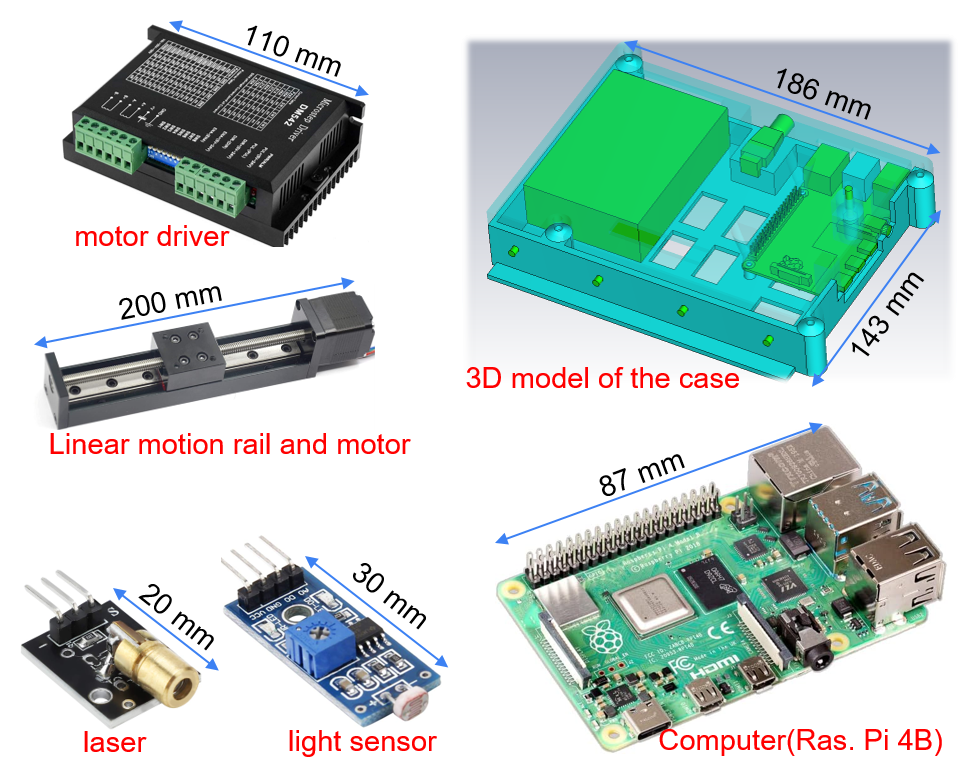}
    \caption{Overview of required components.}
    \label{fig:components}
\end{figure}

\begin{figure}[H]
     \centering
     \begin{subfigure}[a]{0.45\textwidth}
         \centering
         \includegraphics[width=\textwidth]{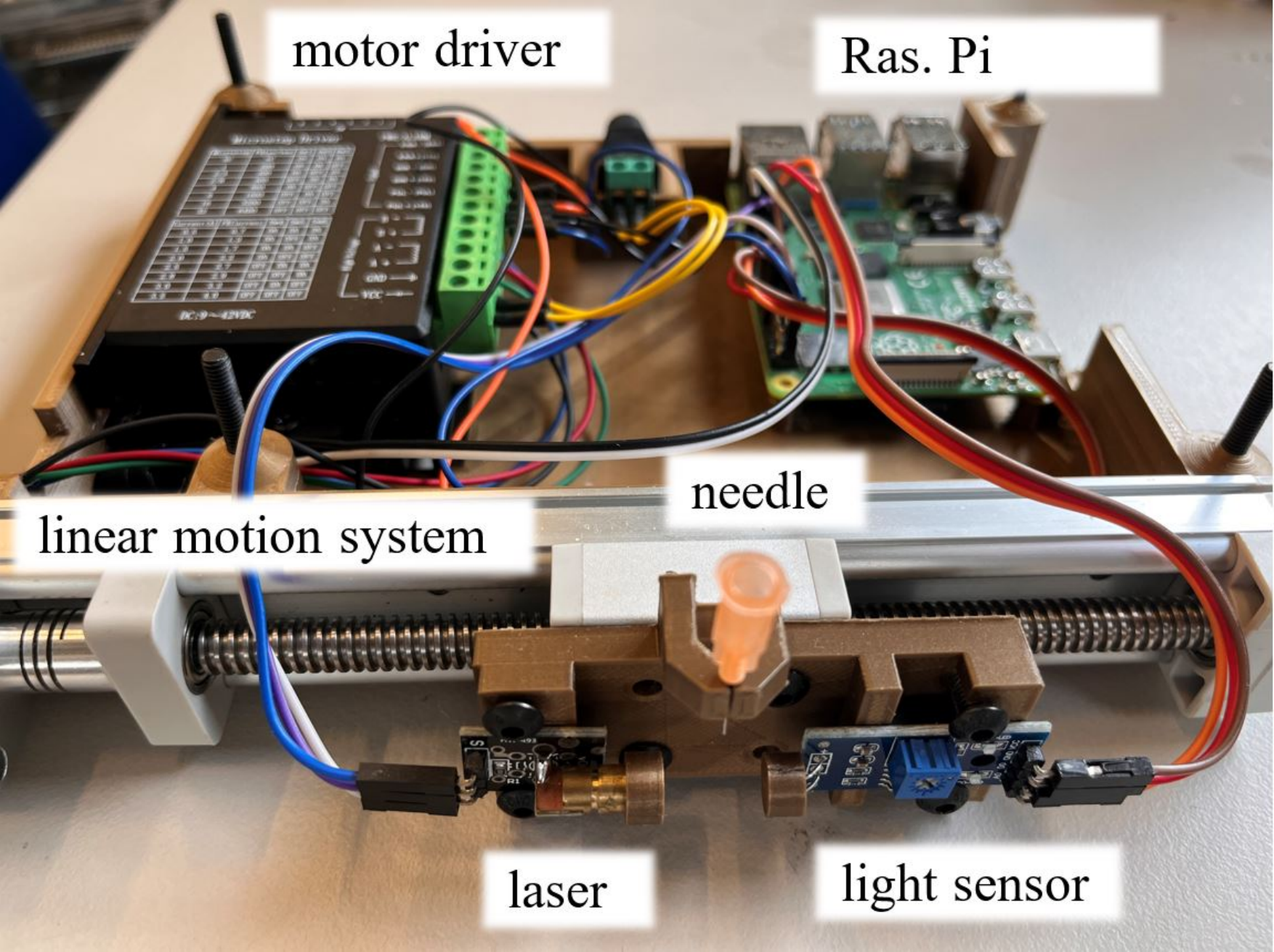}
         \caption{Inner structure overview of the device.}
         \label{fig:a}
     \end{subfigure}
     \hfill
     \begin{subfigure}[b]{0.45\textwidth}
         \centering
         \includegraphics[width=\textwidth]{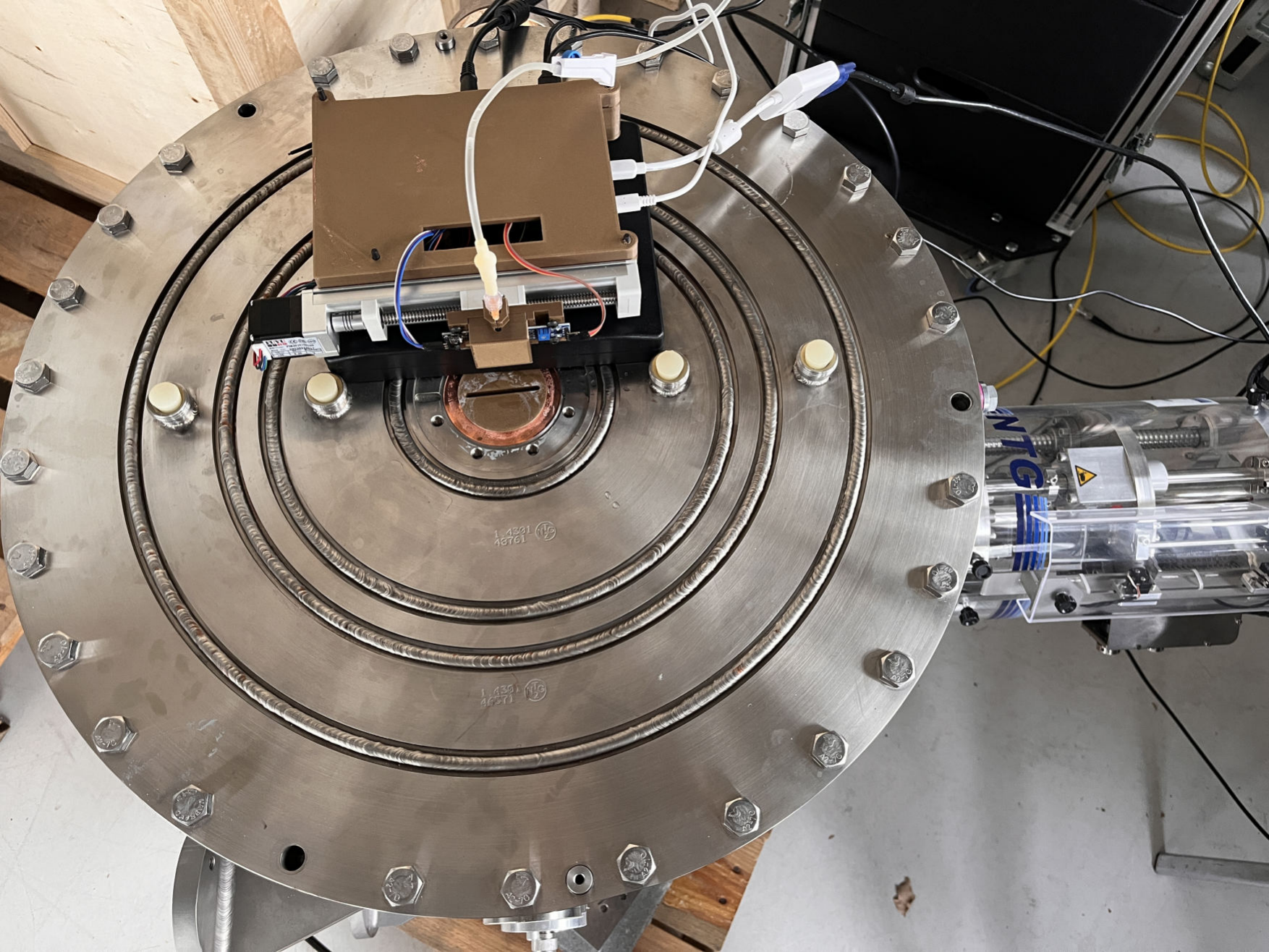}
         \caption{The device is placed on top of the cavity to be measured, connected to a VNA and a monitor.}
         \label{fig:three sin x}
     \end{subfigure}
       \caption{All components are fixed to the 3D printed case~(a). Measurement device installed on top of the cavity to be measured~(b).}
        \label{fig:device}
\end{figure}

After continuous exploration, experimentation, and iterations, the design of the first measurement device has been finalized, which utilizes a series of components as shown in Fig.~\ref{fig:components}. These components are easily accessible, inexpensive, and mature products.
Figure~\ref{fig:device} showcases the device formed after integrating all the components. It is evident that the entire device's volume is only half the size of a letter-sized sheet of paper, making it portable and lightweight compared to the traditional wire-pulley system, and it is universal for all kinds of cavities, except the cavity is not suitable for vertical standing. For the issue of alignment, the current solution is to use a 3D-printed cap (designed according to the cavity) to position the bead. The positioning accuracy depends on the precision of 3D printing, being ensured for typical measurements. The following subsection provides detailed decriptions of each hardware component individually.

\subsection{bead release system}
The perturbator needs to be released with the release point precisely positioned relatively to the chamber (cavity) being tested in both, height and horizontal position. When releasing the perturbator, it is crucial to ensure that it does not acquire any horizontal velocity that could affect its falling path, nor should it have any random initial vertical velocity that could cause deviations in the time-position relationship. For the two types of perturbators, solid beads and liquid droplets, different release mechanisms should be designed. As a starting point, a medical drip infuser has been used to generate continuous droplets of water as perturbators. The blunt tip syringe needles of the infuser can smoothly produce water droplets in a manner that meets the requirements, with the size of the droplets controllable by selecting the size of the needle, and the rate of droplet formation easily regulated using a valve. Additionally, we have designed and 3D printed a fixture part to conveniently secure the needle to the exterior of the device.

\subsection{linear motion system}
The current design utilizes a linear motion system composed of a roller screw to achieve one-dimensional movement. If necessary, this system can be expanded to a two-dimensional motion system. In some cases, where only the electromagnetic field along a single path of the beam axis needs to be measured, the linear motion system can be omitted.

\subsection{bead detection system}
A laser diode and a light sensor are positioned directly opposite to each other and are separated by about two centimeters. They form a photoelectric barrier, refered to as trigger location. The setup is arranged such that the laser diode emits a beam that travels across the gap and is received by the light sensor. Without bead, the light sensor outputs a high voltage level indicating the uninterrupted reception of the laser beam.

The bead is released about 2~cm from above the gap. Prior to the bead reaching the gap, a digital light intensity sensor module, which can switch between Binary Output (BO) and Analog Output (AO), continuously outputs~"1" due to the unobstructed laser light. When the bead reaches the gap and obstructs the laser beam, the amount of light reaching the light sensor decreases below a predefined threshold, causing it to output~"0".

The Raspberry Pi is programmed to monitor the output from the light sensor in real-time. Upon detecting a transition to a low voltage level, indicative of the bead obstructing the laser beam and thus being at the gap position, the Raspberry Pi triggers a VNA also connected to it to commence measurements. This setup allows for precise timing and initiation of measurements based on the physical interruption of a laser beam by a falling object. The total cost of the hardware mentioned so far is below$~300~US\$$.

\subsection{computer (Raspberry PI 4B)}
The system employs the compact Raspberry~Pi~4B (RPI), both as a controller for the VNA and as a computing unit capable of running software for data collection, processing, and visualization. Thanks to its small size (85~mm$\times$56~mm), the Raspberry~Pi can easily be integrated into a portable device. 

\subsection{Vector Network Analyzer}
A crucial point to note during measurements is that the VNA should have a sufficiently high (Intermediate frequencies) IF bandwidth. To avoid undersampling issues, it is necessary to collect enough data points during the short time the perturbator falls through the cavity, especially since the falling speed increases towards the rear end of the cavity, resulting in fewer data points being allocated. The experiments utilized the Rohde $\&$ Schwarz ZNB-4 VNA, which has a maximum IF bandwidth of~10,000, ensuring adequate sampling. For situations where the falling speed is exceptionally fast or the VNA's IF bandwidth is low, leading to undersampling, an alternative solution is available. This involves repeating the falling measurements multiple times, each time initiating the VNA sweep with a slightly different delay to offset the sampling points. Combining the data points from these multiple measurements can resolve the issue of undersampling.

\section{software}
\begin{figure}[H]
    \centering
    \includegraphics[width=8.5 cm]{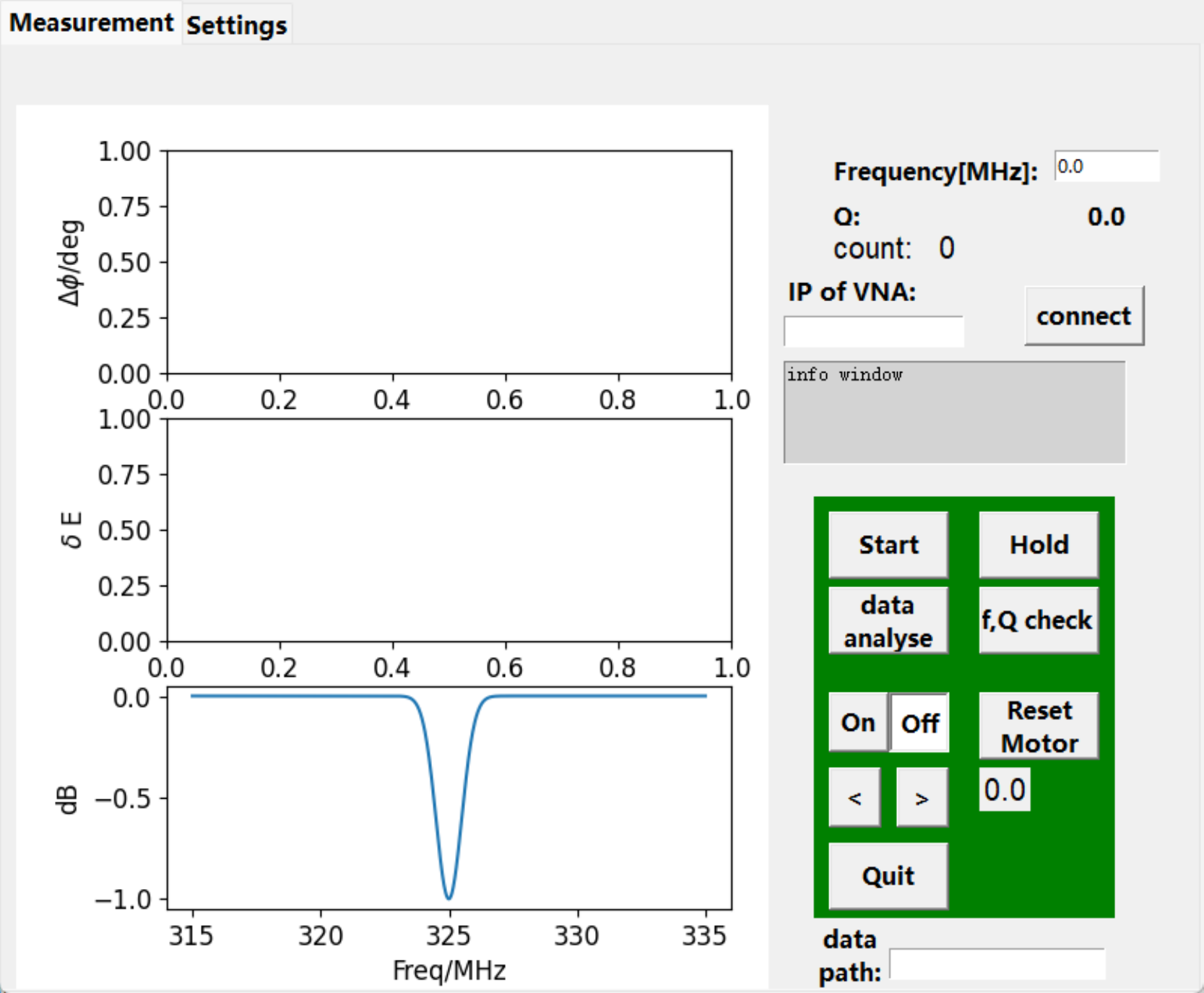}
    \caption{The user interface of the software.}
    \label{fig:UI}
\end{figure}

The software component of the newly developed device is designed to facilitate a seamless interaction between the user and the measurement content and process through its user interface (UI), as detailed in~Fig.~\ref{fig:UI}. This software is pivotal in orchestrating the device's operations, offering several key functionalities that enhance the measurement capabilities and data processing efficiency.
Firstly, the software allows users to customize measurements via input through the UI, tailoring the process to meet specific requirements. It engages in communication with a VNA to dispatch measurement instructions and retrieve data, thereby streamlining the measurement workflow. A significant feature includes the conversion of phase difference-time curves into electric field strength-position curves, which are then visualized through plots to present the measurement outcomes comprehensively.

Additionally, the software is capable of monitoring optical detectors to detect signals triggered by the bead passing through trigger points, subsequently prompting the VNA to initiate a measurement instantly. It also controls the linear motion system to adjust the bead release position, ensuring precise manipulation of the measurement environment.

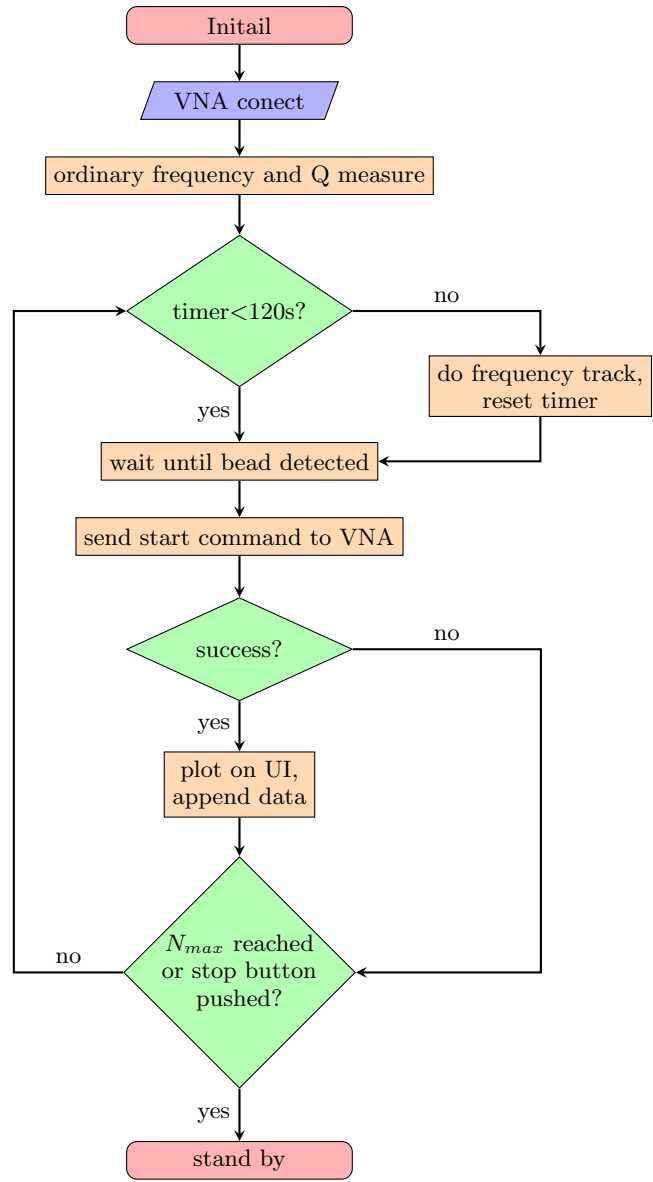
\begin{figure}
    \centering
\begin{tikzpicture}[node distance=2cm]
\node (start) [startstop] {Initail};
\node (in1) [io, below of=start,yshift=+1cm] {VNA conect};
\node (pro1) [process, below of=in1,yshift=+1cm] {ordinary frequency and Q measure};
\node (dec1) [decision, below of=pro1, yshift=+0.2cm] {timer$<$120s?};
\node (pro2a) [process, below of=dec1, yshift=+0cm] {wait until bead detected};
\node (pro2b) [process, right of=dec1, xshift=2cm,align=center,yshift=-1cm] {do frequency track,\\ reset timer};
\node (pro3) [process, below of=pro2a, yshift=1cm] {send start command to VNA};
\node (dec2) [decision, below of=pro3, yshift=0.5cm] {success?};
\node (pro4a) [process, below of=dec2, yshift=0.2cm,align=center] {plot on UI, \\append data};
\node (dec3) [decision, below of=pro4a,yshift=-0.5cm,align=center] {$N_{max}$ reached\\ or stop button\\ pushed?};
\node (stop) [startstop, below of=dec3,yshift=-0.5cm] {stand by};

\draw [arrow] (start) -- (in1);
\draw [arrow] (in1) -- (pro1);
\draw [arrow] (pro1) -- (dec1);
\draw [arrow] (dec1) -- node[anchor=east] {yes} (pro2a);
\draw [arrow] (dec1) -| node[anchor=south,near start] {no} (pro2b);
\draw [arrow] (pro2a) -- (pro3);
\draw [arrow] (pro2b) |- (pro2a);
\draw [arrow] (pro3) -- (dec2);
\draw [arrow] (dec2) -- node[anchor=east] {yes} (pro4a);
\draw [arrow] (dec2.east) -| node[near start, above] {no} ++(2.5,0) |- (dec3.east);
\draw [arrow] (pro4a) -- (dec3);
\draw [arrow] (dec3) -| node[near start, above] {no} ++(-3,0) |- (dec1);
\draw [arrow] (dec3) -- node[anchor=east] {yes} (stop);
\end{tikzpicture}
    \caption{Flow chart of the measuring procedure.}
    \label{fig:flowchart}
\end{figure}

A focal point of the software's design is its ability to automate repeated measurements and process the data efficiently. This capability is critical for rapidly acquiring a large volume of data within a short timeframe, enabling the use of Singular Value Decomposition (SVD) for noise filtering. Such an approach not only significantly enhances the quality of the measured electric field distribution but also allows for noise analysis. The software's automation feature facilitates continuous droplet release, inducing perturbations that enable real-time updates of the field distribution measurements. This is particularly beneficial for the tuning process of cavities, where the ability to quickly adjust and measure changes is crucial. The flow chart of the software is shown in~Fig.~\ref{fig:flowchart}.

Moreover, the software incorporates additional considerations to ensure measurement accuracy and reliability. It is programmed to re-measure the resonance frequency at set intervals (e.g., every 120 seconds) to mitigate the impact of frequency drift on the results. It also includes mechanisms to identify and discard measurements that are compromised due to delays or timeouts in commands sent to the VNA, which could render the results unusable. This level of meticulousness in design and functionality underscores the software's integral role in enhancing the device's performance and reliability in complex measurement scenarios.

\section{measurements on a buncher cavity}
\begin{figure}[H]
	\centering
	\includegraphics[width=8 cm]{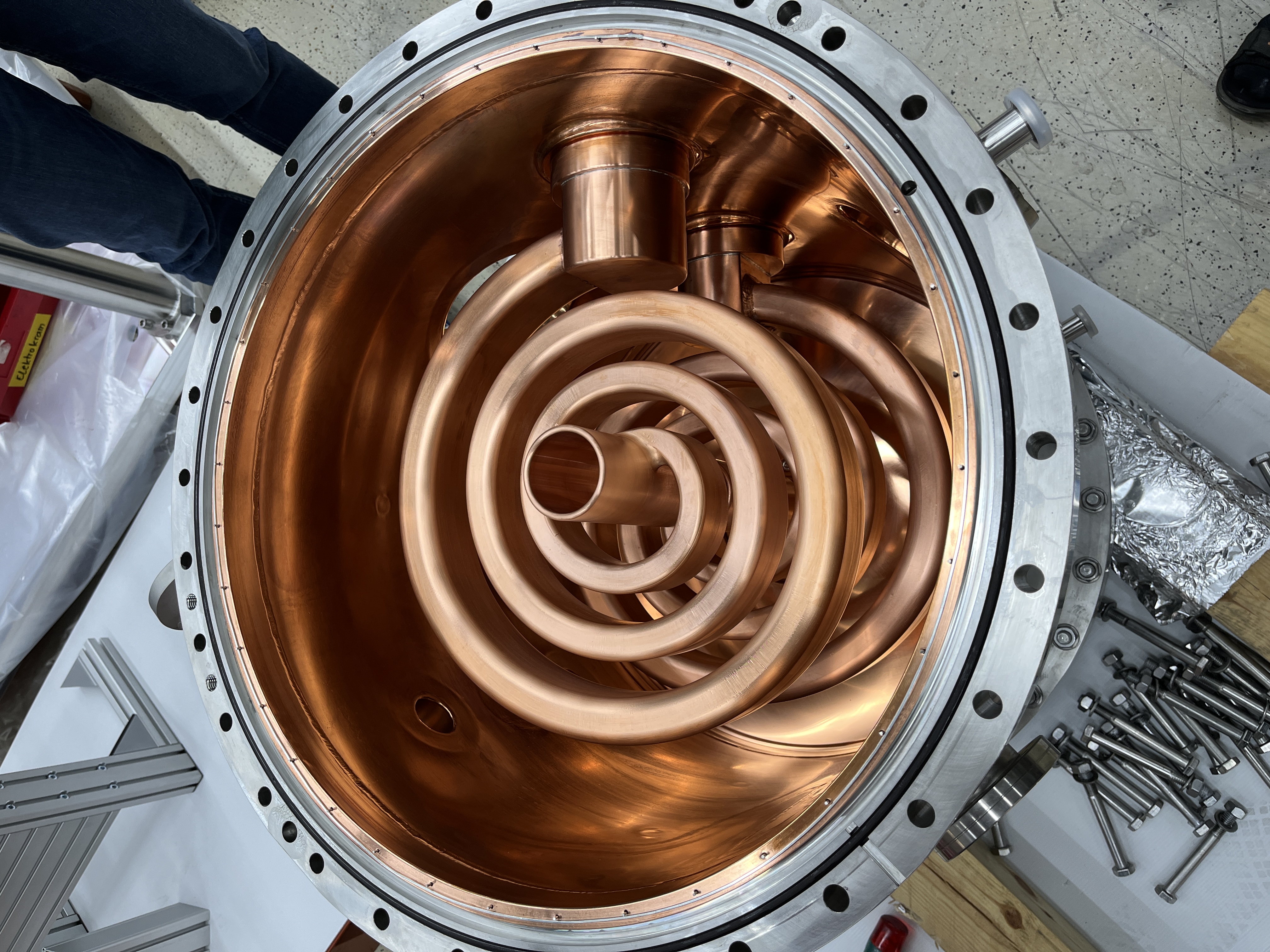}
	\caption{Three-gap buncher cavity of spiral-type, to be operated at 36 MHz.}
	\label{fig:foto_BB3}
\end{figure}

A first set of measurements has been done using a three-gap buncher cavity intended to be installed into an existing accelerator section. It is depicted in Fig.~\ref{fig:foto_BB3} and its parameters are listed in~Table~\ref{Tab:BB3para}.
 
\begin{table}[H]
	\label{tab_BB3}
\centering
\caption{Parameters of the buncher BB3}
\begin{tabular}{|c|c|c|}
\hline
\textbf{Parameter} & \textbf{Unit} & \textbf{Value} \\
\hline
RF-frequency & MHz & 36.136 \\
Gaps & \# & 3 \\
Gap length & mm & 12.7–13.5 \\
Drift tubes & \# & 2 \\
Drift tube length & mm & 38.1–40.8 \\
Aperture & mm & 100 \\
Tank diameter & mm & 420 \\
Tank length & mm & 525.7 \\
Q-factor & & 1000 \\
\hline
\end{tabular}
\label{Tab:BB3para}
\end{table}

The cavity's orientation aligns the beam axis w.r.t.~local gravity, with the measurement apparatus mounted atop the cavity along the beam axis. Due to low Q value~(1000) of this buncher with long spirals, in the measured data, the magnitude of noise in the measurement data is close to the magnitude of the signal itself. Though fast iterative measurements allow to collect big mount of data within short time.

Before the device is put into application, two fixed parameters need to be determined. The first one is the speed~$v_0$ of the bead when passing the trigger location. The second parameter to be determined is the system delay~$t_d$, designated as, which is the time delay between the bead passing the trigger location and the VNA generating the first measurement point. A cavity with three equidistant gaps is suitable for calibrating these two parameters. The distance between the three gaps of the cavity is known and taken as a constant~$L$. Conducting a measurement will produce a curve with three distinct peaks corresponding to the three gaps. The speed of the falling bead at all measurement points can be inferred from the distances between these three peaks, thereby deducing the speed of the bead at the trigger location and hence~$t_d$. The bead is dropped from different heights into the cavity at different speeds such that more reliable determinations of~$v_0$ and~$t_d$ can be obtained by averaging.

\begin{figure}[H]
    \centering
    \includegraphics[width=8.5cm]{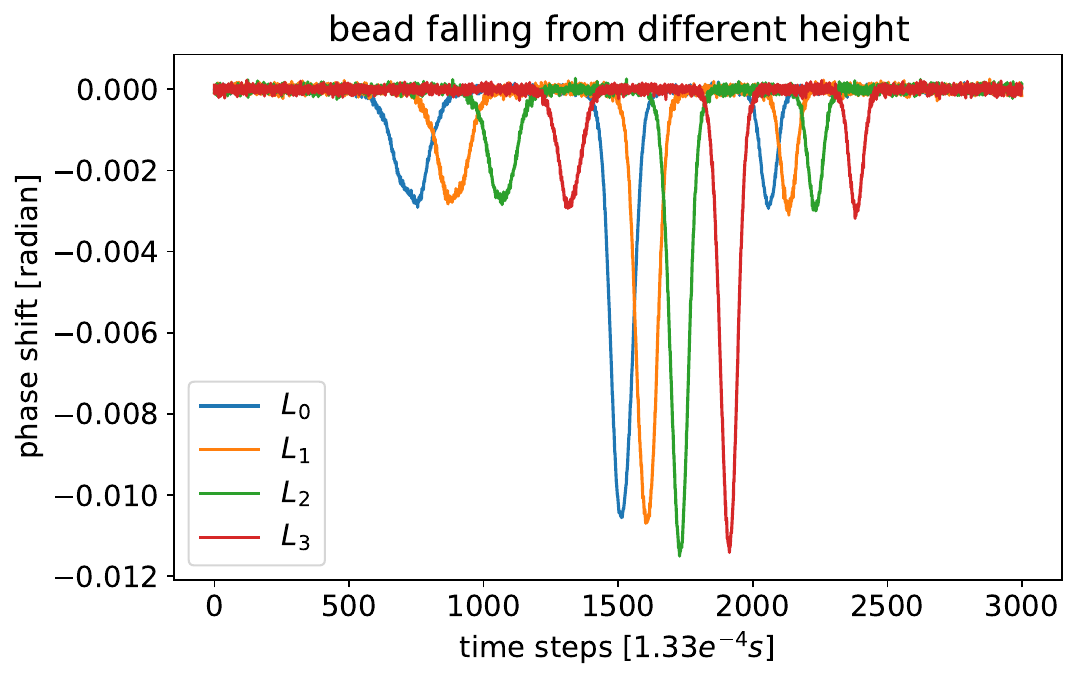}
    \caption{Phase shift data measured from the bead launched from four different heights $L_0$, $L_1$, $L_2$ and $L_3$ into the buncher cavity.}
    \label{fig:phases_4}
\end{figure}

Figure~\ref{fig:phases_4} depicts an example for such measurements. After these two parameters have been determined, the device is ready for measurements on any other type of cavities without needing to repeat the aforementioned process. In the bead falling measurement, the distance between the device and the cavity entrance is a known constant. Therefore, the measured time-phase shift curves can be translated into position-field strength curves.

\begin{figure}[H]
    \centering
    \centering
    \includegraphics[width=8.5 cm]{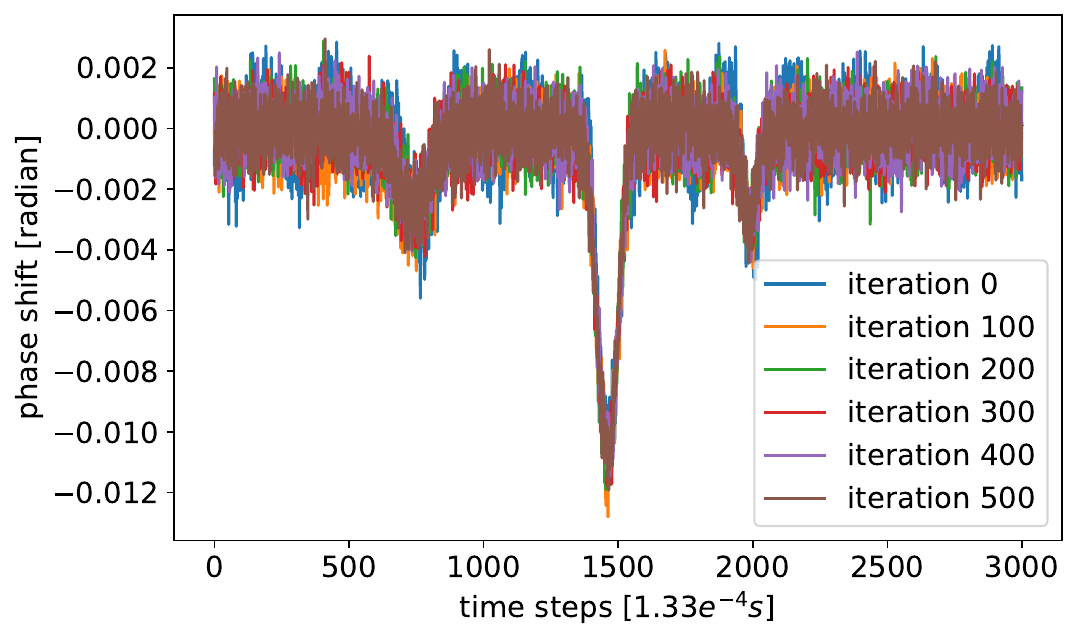}
    \caption{Measured data taken with the buncher cavity with 3000 discrete time points within 0.4 second each, corresponding to a length of about 1.2~m along the cavity axis.}
    \label{fig:BB3dataCheck_1d}
\end{figure}

The noisy data set obtained from the perturbation method is compiled into 1D arrays, making it suitable for processing with SVD. It should be noted that SVD is performed only after multiple iterations and sufficient data collection; it does not play a role in the real-time field distribution monitoring at 2~Hz mentioned earlier. Therefore, to remove noise from the updates of real-time field distribution, other algorithms should be considered. However, before applying SVD, the measured frequency variation data must be converted into phase shift data by subtracting it with a reference phase. The reference phase for each iteration is determined by taking the average phase of the points near both boundaries. Each iteration provides a 1D array of 3000 points corresponding to time steps. After 500 iterations, a 2D data array ($500\times3000$) is obtained. Figure~\ref{fig:BB3dataCheck_1d} shows a summarizing plot.

\begin{figure}[H]
    \centering
    \centering
    \includegraphics[width=8.5cm]{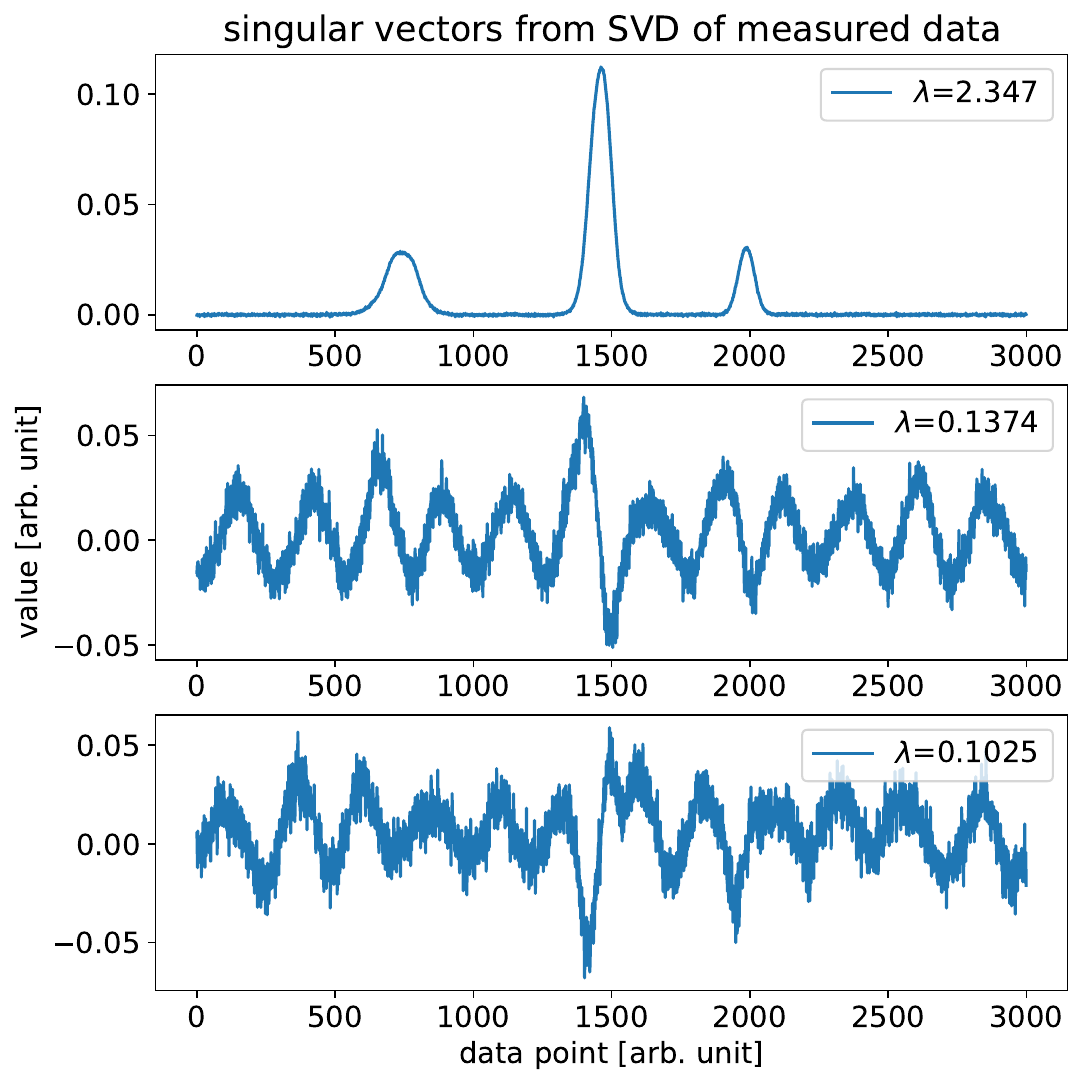}
    \caption{First three singular vectors obtained from SVD (buncher cavity), ordered by the magnitude of their corresponding singular values.}
    \label{fig:BB3datasvd}
\end{figure}

After applying SVD (see Appendix B), the data is decomposed into several components, with the first three singular vectors shown in~Fig.~\ref{fig:BB3datasvd}. The first singular vector is essentially considered to be the noise-free essence of the data. It is treated and compared with simulated field distributions.

\begin{figure}[H]
    \centering
    \includegraphics[width=8.5cm]{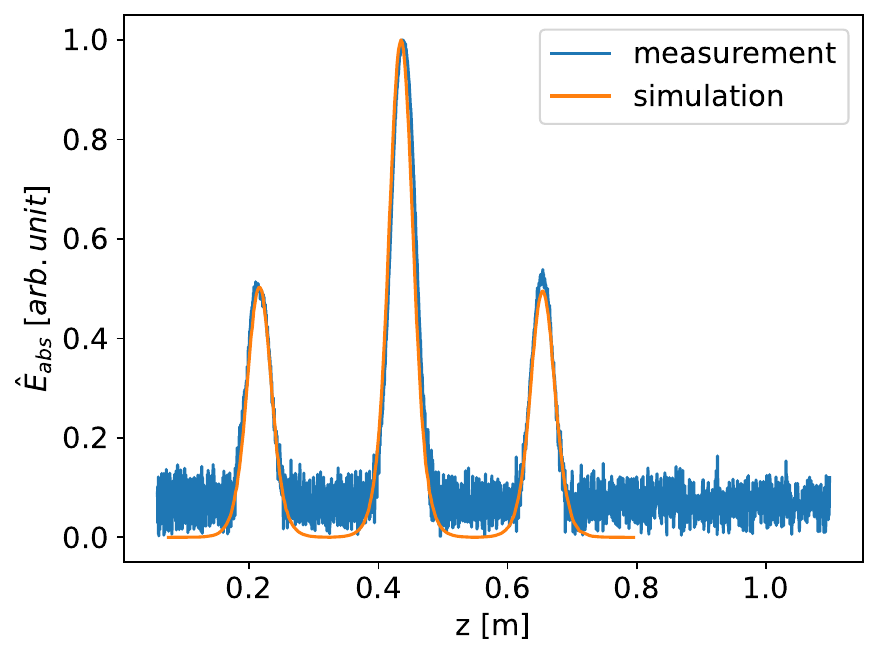}
    \caption{Measured electric field(denoted by $\hat{E}_{abs}$ along the axis of the buncher cavity compared with simulation.}
    \label{fig:BB3_compare}
\end{figure}

Comparison between measured field and Micro-Wave-Studio (MWS) simulations~\cite{CSTStudioSuite2023} is shown in~Fig.~\ref{fig:BB3_compare}.
There are some deviations as a result of the finite accuracies of the machining, eigenmode position definition, and the measurement itself.

\begin{figure}[hbt]
        \centering
        \includegraphics[width=8.5cm]{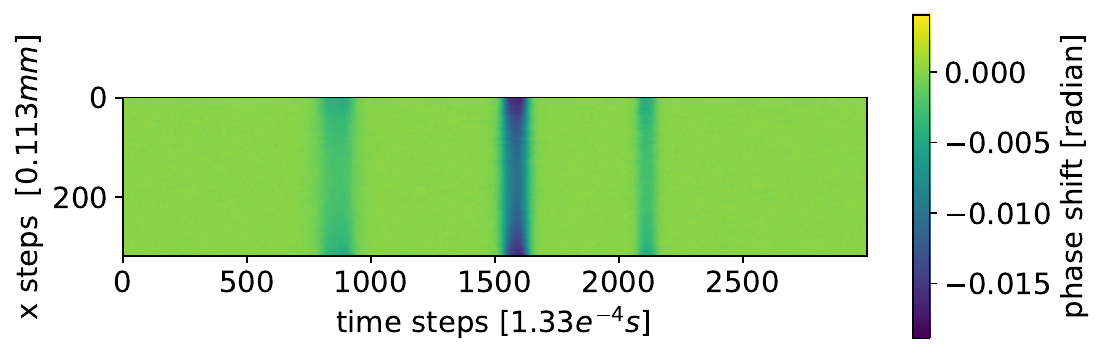}
        \caption{Original 2D array data from the buncher cavity with even time interval of~$dt=0.4s/3000$.}
        \label{fig:2D_data_t}
\end{figure}

Afterwards, 2D field measurements have been performed on the same cavity. The positioning of the bead is achieved using a 3D-printed cap that fits the entrance of the cavity, with a slit constraining the position of the droplets. The length of the slit covers the off-axis area intended for measurement, and its width is just sufficient to allow a water droplet to pass through. In each iteration, the needle is moved a small constant distance, and a 1D field along the path of free falling is measured. As the needle moves from one end of the slit to the other, the measurement of the entire 2D electric field is completed. Figure~\ref{fig:2D_data_t} show the measured 2D array data in the form of a bitmap. This data is the phase shift as a function of time, measured at different off-axis distances. it is converted to 2D field distribution, according to time-height relation presented in appendix~A.

\begin{figure}[H]
    \centering
    \includegraphics[width=8.5 cm]{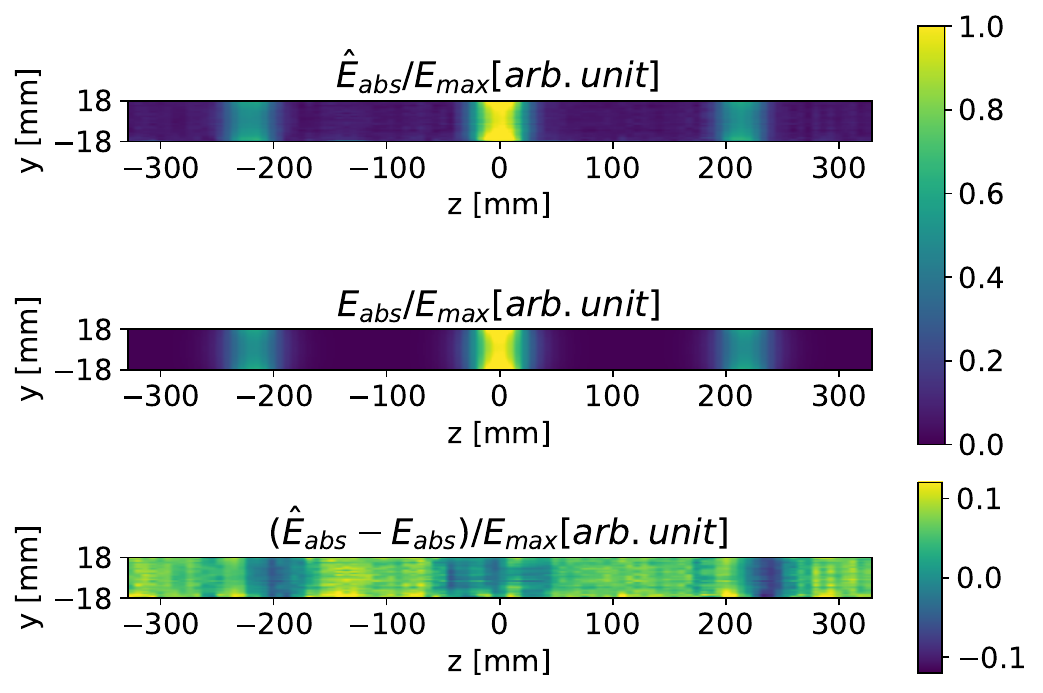}
    \caption{The distribution $\hat{E}_{abs}$ represents the experimentally measured two-dimensional field, while $E_{abs}$ denotes the corresponding field obtained from simulations. Both fields are normalized to their respective maxima, specifically the peak value observed at the middle gap. The term $\hat{E}_{abs}-E_{abs}$ quantifies the deviation of the measured values from the simulated ones.}
    \label{fig:2d_compare}
\end{figure}

Then, with proper translation and scaling, the measured field(denoted by $\hat{E}_{abs}$) can be compared with corresponding field obtained from simulation(denoted by $E_{abs}$), as shown in Figure~\ref{fig:2d_compare}.

\section{measurement on a 1:3 aluminum model cavity of Alvarez type}
\begin{figure}[H]
    \centering
    \includegraphics[width=8 cm]{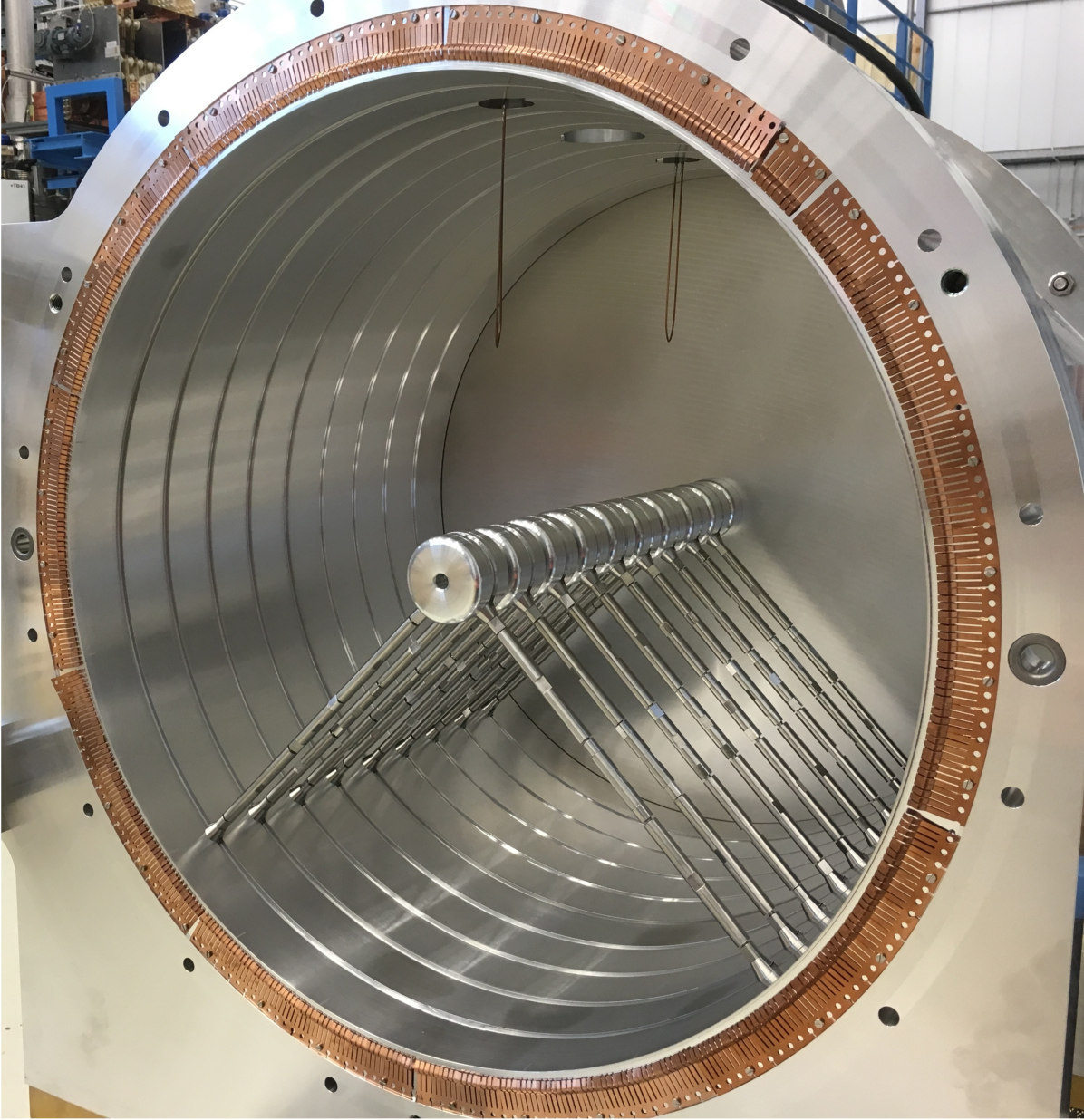}
    \caption{Scaled (1:3) aluminum model of an Alvarez-type cavity.}
    \label{fig:model_cavity}
\end{figure}

The measurement process was similarly applied to a model cavity, specifically a 1:3 scale aluminum replica of a 10-gap Alvarez-type cavity. This model primarily served to explore an efficient stabilization strategy, focusing on tilt sensitivity~\cite{Seibel2017,PhysRevAccelBeams.20.032001}. It is depicted in~Fig.~\ref{fig:model_cavity} and its parameters are listed in~Table~\ref{fig:model_cavity}.

\begin{table}[H]
\centering
\label{tab_model}
\caption{Parameters of the scaled (1:3) Alvarez-type cavity model.}
\begin{tabular}{|c|c|c|}
\hline
\textbf{Parameter} & \textbf{Unit} & \textbf{Value} \\
\hline
RF-frequency & MHz & 325.224 \\
Gaps & \# & 10 \\
Gap length & mm & 12.7–13.5 \\
Drift tubes & \# & 9 \\
Drift tube length & mm & 38.1–40.8 \\
Drift tube diameter & mm & 60.0 \\
Aperture & mm & 10.0 \\
Tank diameter & mm & 634.5 \\
Tank length & mm & 525.7 \\
Q-factor & & 500 \\
\hline
\end{tabular}
\end{table}

\begin{figure}[H]
    \centering
    \includegraphics[width=9cm]{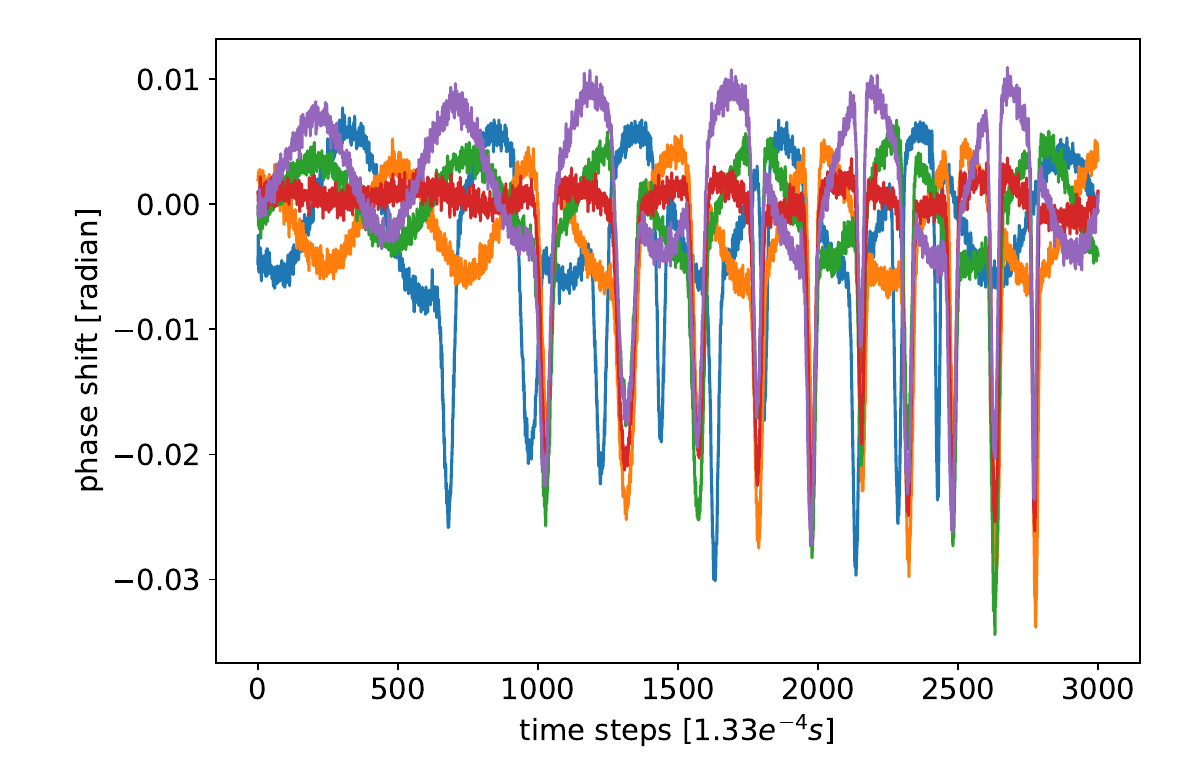}
    \caption{Measured phase shift data from the DTL model cavity, with a signal-to-noise ratio of approximately~2. Different colors represent multiple phase shift measurements conducted using the DTL. Only five representative results from thousands are presented. }
    \label{fig:model_noisy}
\end{figure}

A phase shift measurement was chosen rather than resonant frequency shift as the phase is signiﬁcantly easier to detect. The measurement exhibited a very low signal-to-noise ratio, approaching 2:1, due to three main reasons: 1) The cavity's Q-value of~530 was low because it was made from aluminum. 2) The testing environment included background noise from surrounding equipment, such as vacuum pumps, contributing to the overall noise level. 3) The coupling loop and pickup were made of loose copper wires, which, under the influence of background mechanical vibrations, created relative vibrations with the cavity. These adverse factors are usually strenuously avoided in conventional bead pull measurements. However, our bead-fall method and the corresponding device made it possible to conduct measurements in such an extremely unfavorable environment. Figure~\ref{fig:model_noisy} plots the measured phase shifts.

\begin{figure}[H]
    \centering
    \includegraphics[width=9cm]{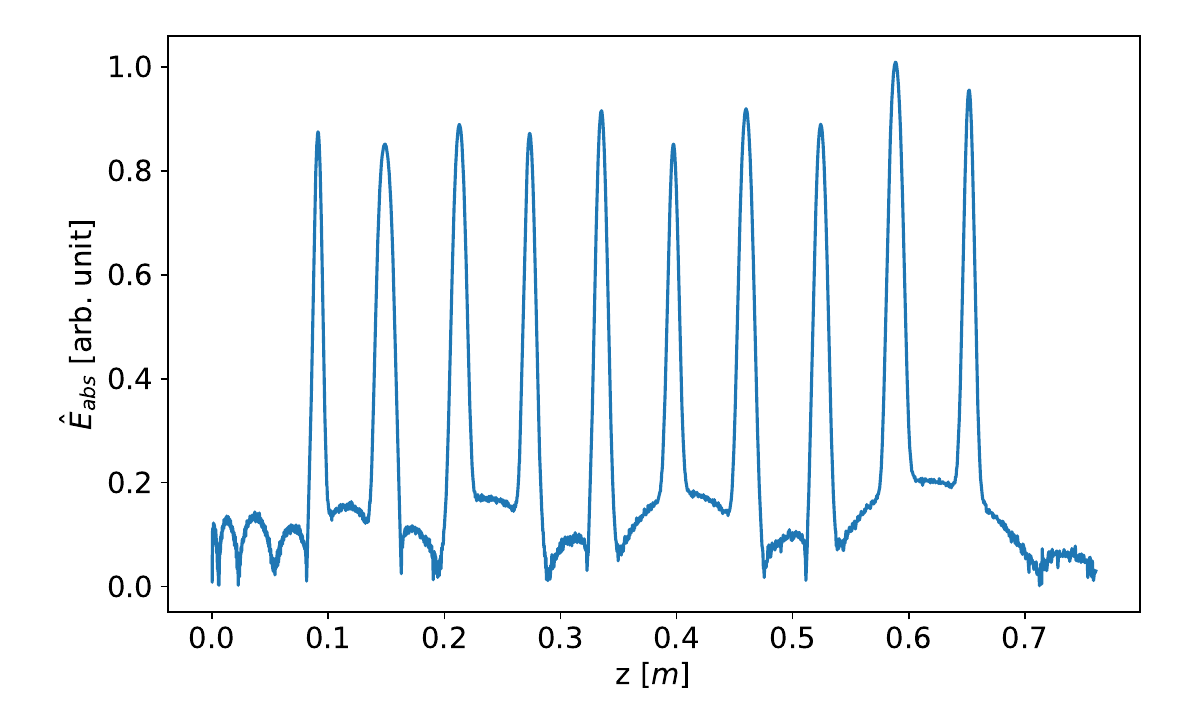}
    \caption{Measured electric field along the beam axis of the DTL model cavity.}
    \label{fig:model_field}
\end{figure}

Despite the initial data being heavily laden with noise, attributed to the soft coupler, we successfully discerned the field distribution. The ability to perform rapid, iterative measurements facilitated the accumulation of a substantial dataset in a brief timeframe. This wealth of data enabled us to employ SVD techniques to filter out noise and accurately ascertain the field distribution, as depicted in Fig.~\ref{fig:model_field}.

Alvarez-type cavities are characterized by heavy drift tubes suspended by support rods from the upper side of a horizontally positioned cavity cylinder. However, during measurements, the cavity is placed vertically, leading to gravity causing the support rods to bend downwards, which in turn leads to unpredictable changes in the distance between the drift tube gaps. Additionally, the cavity was not precisely aligned. These factors resulted in a measured field distribution that could not be directly compared with theoretical values. Nonetheless, the measured field distribution clearly displayed the gap fields of ten cells, providing a reliable basis for tuning the field flatness.

\section{conclusion}
This study introduces an innovative bead-falling measurement technique, establishing a groundbreaking alternative to traditional RF-field mapping methods, particularly the bead-pull approach. The technique is underscored by its streamlined operational efficiency, reducing mechanical complexity and facilitating rapid, successive measurements. Employing SVD for meticulous noise filtration, this method delineates precise field distributions from noisy data, demonstrating exceptional accuracy.

A key highlight of our research is the development of a new measurement device characterized by its compactness, portability, and universal applicability across a diverse range of cavity types. This device, integral to the bead-falling measurement technique, has been rigorously tested under challenging conditions, including low signal-to-noise ratios and environmental vibrations. Its robust design and operational resilience were evident in the accurate discernment of field distributions in both a three-gap buncher cavity and a 10-gap Alvarez-type cavity model, confirming its efficacy in adverse settings.

The device's minimalist design simplifies the preparatory phase of experiments, significantly reducing setup times and facilitating swift measurement cycles. This advancement not only enhances the practicality of field distribution measurements but also extends the scope of application to various cavity configurations without the need for extensive modifications or customization. The measurement process's efficiency, coupled with the device's versatility, underscores the potential for widespread adoption in accelerator physics and resonant cavity diagnostics.

Looking ahead, we aim to further refine this technique and its associated hardware, broadening its application spectrum and leveraging its capabilities to optimize cavity diagnostics and tuning. The introduction of the bead-falling measurement method and its accompanying device offers a potent tool for the precise evaluation and enhancement of electromagnetic field distributions in resonant cavities.

\appendix

\section{consideration of air resistance during free fall}

A spherical bead with the cross section area~$A$ which falls at velocity of~$|v|$ experiences the friction force from air having the density~$\rho$=1.2~kg/m$^{\text{3}}$ of
\begin{equation}
F_{air}=\frac{1}{2}\rho AC_d |\vec{v}^2|\,,
\end{equation}
where~$C_d$ is the dimensionless drag coefficient being equal to~0.47 for spheres. The falling speed as a function of time reads as
\begin{equation}
    v(t)=\sqrt{\frac{2mg}{\rho A C_d}} \tanh{(t \sqrt{\frac{g\rho C_d A}{2m}})}\,.
\end{equation}

Defining $k:=\frac{1}{2}\rho AC_d$, the height as a function of time turns into
\begin{equation}
h(t)\,=\, h_0\,+\,\frac{m \ln{(\cosh{\sqrt{\frac{gk}{m}}t})}}{k}
\end{equation}
where~$h_0$ is the initial height, $m$ is the spheres mass, and~$g$ is the acceleration from gravity.
With advancing time, the velovity converges towards the stable velocity of
\begin{equation}
v_s=\sqrt{\frac{2mg}{\rho C_{D}A}}.   
\end{equation}
Within relevant heights of few meters, translations from the coriolis force can be neglected.

\section{Singular value decomposition in data process}

SVD, or Singular Value Decomposition, is a mathematical technique used in linear algebra to decompose a matrix into three distinct matrices, specifically designed to identify the intrinsic geometric structure of the data. The decomposition is expressed as:
\begin{equation}
    A=USV^T\,,
\end{equation}
where $A$ is the original matrix, $U$ is an $m\times m$ orthogonal matrix whose columns are the left singular vectors of $A$,
$S$ is an $m\times n$ rectangular diagonal matrix with non-negative real numbers on the diagonal, known as the singular values of $A$, $V^T$ (the transpose of 
V) is an $n\times n$ orthogonal matrix whose columns are the right singular vectors of $A$.
The singular values in are typically arranged in descending order. This ordering is crucial because the first few singular values (and their corresponding singular vectors) often capture the most significant modes of variation in the data, whereas the smaller singular values tend to correspond to noise or less important details.

One of the most significant advantages of SVD in the context of noisy data is its ability to perform noise reduction or data smoothing. By truncating the $S$ matrix to retain only the largest singular values (and ignoring the smaller ones that often correspond to noise), SVD can reconstruct a version of the original matrix that preserves the most critical structures while filtering out noise. This process is known as low-rank approximation.

SVD is particularly useful in signal processing and statistics for noise reduction. By identifying the singular values, which are essentially the strengths of the signals within the data, it allows for the separation of the "signal" (important information) from the "noise" (unimportant information). This is because the singular values corresponding to noise tend to be significantly smaller than those corresponding to the signal. By truncating or ignoring the smaller singular values (and their corresponding vectors), we can approximate the original data with less noise. Details including successfull application of SVD to simulated RF-cavity fields can be found in \cite{PhysRevAccelBeams.21.084601}.

\section{Coriolis force}
The Coriolis force is an apparent force that arises from the rotation of the Earth. It is an inertial force described mathematically by the Coriolis effect. To understand it intuitively, consider that the Earth rotates from west to east; this rotation influences the movement of air, water, and even objects in motion across its surface.

When an object falls freely (like the bead or droplet dropped from a height), it not only moves downward due to gravity but also moves horizontally due to its initial horizontal component of velocity (if any) and due to Earth’s rotation. Here's how the Coriolis force affects it:

The magnitude of this deflection depends on several factors:
The height from which the object falls (greater height means more time to be deflected).
The latitude (effect is maximal at the poles and zero at the equator).
The object's speed of descent.

Consider a freely falling object at a latitude \(\phi\). The Coriolis acceleration \(a_c\) can be approximated by the formula:
\[
a_c = 2\Omega v \sin(\phi)\,
\]
where:
 $\Omega \approx$7.27 $\cdot$ 10$^\text{-5}$ rad/s is the Earth's angular velocity, $v$is the velocity of the falling object, and~$\phi$ is the latitude at which the object is falling.

For an object starting from rest and falling a distance~$h$, the time~$t$ of fall can be calculated using:
\[
t = \sqrt{\frac{2h}{g}}\,.
\]

The acceleration \(a_c\) leads to a deflection which can be integrated over time to find the eastward displacement due to the Coriolis force alone. This deflection is given by
\[
d = \frac{1}{2} a_c t^2\,.
\]
Considering the latitude~$\phi$ at the GSI laboratory of~50$^\text{o}$ and assuming a launching height of~2~m, the eastward displacement is less than~0.15~mm, being negligible.

\nocite{*}

\bibliography{apssamp}

\providecommand{\noopsort}[1]{}\providecommand{\singleletter}[1]{#1}%
\begin{thebibliography}{17}%
\makeatletter
\providecommand \@ifxundefined [1]{%
 \@ifx{#1\undefined}
}%
\providecommand \@ifnum [1]{%
 \ifnum #1\expandafter \@firstoftwo
 \else \expandafter \@secondoftwo
 \fi
}%
\providecommand \@ifx [1]{%
 \ifx #1\expandafter \@firstoftwo
 \else \expandafter \@secondoftwo
 \fi
}%
\providecommand \natexlab [1]{#1}%
\providecommand \enquote  [1]{``#1''}%
\providecommand \bibnamefont  [1]{#1}%
\providecommand \bibfnamefont [1]{#1}%
\providecommand \citenamefont [1]{#1}%
\providecommand \href@noop [0]{\@secondoftwo}%
\providecommand \href [0]{\begingroup \@sanitize@url \@href}%
\providecommand \@href[1]{\@@startlink{#1}\@@href}%
\providecommand \@@href[1]{\endgroup#1\@@endlink}%
\providecommand \@sanitize@url [0]{\catcode `\\12\catcode `\$12\catcode `\&12\catcode `\#12\catcode `\^12\catcode `\_12\catcode `\%12\relax}%
\providecommand \@@startlink[1]{}%
\providecommand \@@endlink[0]{}%
\providecommand \url  [0]{\begingroup\@sanitize@url \@url }%
\providecommand \@url [1]{\endgroup\@href {#1}{\urlprefix }}%
\providecommand \urlprefix  [0]{URL }%
\providecommand \Eprint [0]{\href }%
\providecommand \doibase [0]{https://doi.org/}%
\providecommand \selectlanguage [0]{\@gobble}%
\providecommand \bibinfo  [0]{\@secondoftwo}%
\providecommand \bibfield  [0]{\@secondoftwo}%
\providecommand \translation [1]{[#1]}%
\providecommand \BibitemOpen [0]{}%
\providecommand \bibitemStop [0]{}%
\providecommand \bibitemNoStop [0]{.\EOS\space}%
\providecommand \EOS [0]{\spacefactor3000\relax}%
\providecommand \BibitemShut  [1]{\csname bibitem#1\endcsname}%
\let\auto@bib@innerbib\@empty
\bibitem [{\citenamefont {Maier}\ and\ \citenamefont {Slater}(1952)}]{Maier1952FieldSM}%
  \BibitemOpen
  \bibfield  {author} {\bibinfo {author} {\bibfnamefont {L.~C.}\ \bibnamefont {Maier}}\ and\ \bibinfo {author} {\bibfnamefont {J.~C.~.}\ \bibnamefont {Slater}},\ }\bibfield  {title} {\bibinfo {title} {Field strength measurements in resonant cavities},\ }\href {https://api.semanticscholar.org/CorpusID:123235264} {\bibfield  {journal} {\bibinfo  {journal} {Journal of Applied Physics}\ }\textbf {\bibinfo {volume} {23}},\ \bibinfo {pages} {68} (\bibinfo {year} {1952})}\BibitemShut {NoStop}%
\bibitem [{\citenamefont {Waldron}(1960)}]{Waldron1960PerturbationTheory}%
  \BibitemOpen
  \bibfield  {author} {\bibinfo {author} {\bibfnamefont {R.}~\bibnamefont {Waldron}},\ }\bibfield  {title} {\bibinfo {title} {Perturbation theory of resonant cavities},\ }\href {https://doi.org/10.1049/pi-c.1960.0041} {\bibfield  {journal} {\bibinfo  {journal} {Proceedings of the IEE - Part C: Monographs}\ }\textbf {\bibinfo {volume} {107}},\ \bibinfo {pages} {272} (\bibinfo {year} {1960})}\BibitemShut {NoStop}%
\bibitem [{\citenamefont {Hahn}\ \emph {et~al.}(2011)\citenamefont {Hahn}, \citenamefont {Xu}, \citenamefont {Jain},\ and\ \citenamefont {Johnson}}]{Hahn2011HOMIdentification}%
  \BibitemOpen
  \bibfield  {author} {\bibinfo {author} {\bibfnamefont {H.}~\bibnamefont {Hahn}}, \bibinfo {author} {\bibfnamefont {W.}~\bibnamefont {Xu}}, \bibinfo {author} {\bibfnamefont {P.}~\bibnamefont {Jain}},\ and\ \bibinfo {author} {\bibfnamefont {E.~C.}\ \bibnamefont {Johnson}},\ }\bibfield  {title} {\bibinfo {title} {Hom identification and bead pulling in the brookhaven erl},\ }in\ \href@noop {} {\emph {\bibinfo {booktitle} {Proceedings of the 2011 SRF Conference, Chicago, IL, USA}}},\ \bibinfo {editor} {edited by\ \bibinfo {editor} {\bibfnamefont {S.}~\bibnamefont {Arora}}}\ (\bibinfo {organization} {Brookhaven National Laboratory},\ \bibinfo {address} {Upton, NY 11973, USA},\ \bibinfo {year} {2011})\ p.\ \bibinfo {pages} {THPO041},\ \bibinfo {note} {supported by Brookhaven Science Associates, LLC under Contract No. DE-AC02-98CH10886 with the U.S. DOE and DOE grant DE-SC0002496 to Stony Brook University}\BibitemShut {NoStop}%
\bibitem [{\citenamefont {Koubek}\ \emph {et~al.}(2017)\citenamefont {Koubek}, \citenamefont {Grudiev},\ and\ \citenamefont {Timmins}}]{PhysRevAccelBeams.20.080102}%
  \BibitemOpen
  \bibfield  {author} {\bibinfo {author} {\bibfnamefont {B.}~\bibnamefont {Koubek}}, \bibinfo {author} {\bibfnamefont {A.}~\bibnamefont {Grudiev}},\ and\ \bibinfo {author} {\bibfnamefont {M.}~\bibnamefont {Timmins}},\ }\bibfield  {title} {\bibinfo {title} {rf measurements and tuning of the 750 mhz radio frequency quadrupole},\ }\href {https://doi.org/10.1103/PhysRevAccelBeams.20.080102} {\bibfield  {journal} {\bibinfo  {journal} {Phys. Rev. Accel. Beams}\ }\textbf {\bibinfo {volume} {20}},\ \bibinfo {pages} {080102} (\bibinfo {year} {2017})}\BibitemShut {NoStop}%
\bibitem [{\citenamefont {Berrutti}\ \emph {et~al.}(2014)\citenamefont {Berrutti}, \citenamefont {Khabiboulline}, \citenamefont {Poloubotko}, \citenamefont {Romanov}, \citenamefont {Steimel}, \citenamefont {Yakovlev}, \citenamefont {Li},\ and\ \citenamefont {Staples}}]{Berrutti2014PXIERFQBeadPull}%
  \BibitemOpen
  \bibfield  {author} {\bibinfo {author} {\bibfnamefont {P.}~\bibnamefont {Berrutti}}, \bibinfo {author} {\bibfnamefont {T.~N.}\ \bibnamefont {Khabiboulline}}, \bibinfo {author} {\bibfnamefont {V.}~\bibnamefont {Poloubotko}}, \bibinfo {author} {\bibfnamefont {G.}~\bibnamefont {Romanov}}, \bibinfo {author} {\bibfnamefont {J.}~\bibnamefont {Steimel}}, \bibinfo {author} {\bibfnamefont {V.}~\bibnamefont {Yakovlev}}, \bibinfo {author} {\bibfnamefont {D.}~\bibnamefont {Li}},\ and\ \bibinfo {author} {\bibfnamefont {J.}~\bibnamefont {Staples}},\ }\bibfield  {title} {\bibinfo {title} {Pxie rfq bead pull measurements},\ }in\ \href {https://accelconf.web.cern.ch/LINAC2014/papers/TUPP047.pdf} {\emph {\bibinfo {booktitle} {Proceedings of the 27th Linear Accelerator Conference (LINAC2014)}}},\ \bibinfo {editor} {edited by\ \bibinfo {editor} {\bibfnamefont {C.}~\bibnamefont {Carli}}}\ (\bibinfo  {publisher} {JACoW},\ \bibinfo {address} {Geneva, Switzerland},\ \bibinfo {year} {2014})\ p.\ \bibinfo {pages}
  {TUPP047}\BibitemShut {NoStop}%
\bibitem [{\citenamefont {Goudket}\ \emph {et~al.}(2008)\citenamefont {Goudket}, \citenamefont {Beard}, \citenamefont {McIntosh}, \citenamefont {Burt}, \citenamefont {Dexter},\ and\ \citenamefont {Jones}}]{Goudket2008Comparison}%
  \BibitemOpen
  \bibfield  {author} {\bibinfo {author} {\bibfnamefont {P.}~\bibnamefont {Goudket}}, \bibinfo {author} {\bibfnamefont {C.}~\bibnamefont {Beard}}, \bibinfo {author} {\bibfnamefont {P.}~\bibnamefont {McIntosh}}, \bibinfo {author} {\bibfnamefont {G.}~\bibnamefont {Burt}}, \bibinfo {author} {\bibfnamefont {A.}~\bibnamefont {Dexter}},\ and\ \bibinfo {author} {\bibfnamefont {R.~M.}\ \bibnamefont {Jones}},\ }\bibfield  {title} {\bibinfo {title} {Comparison of stretched-wire, bead pull and numerical impedance calculations on 3.9 ghz dipole cavities},\ }in\ \href {https://accelconf.web.cern.ch/e08/papers/mopp128.pdf} {\emph {\bibinfo {booktitle} {Proceedings of the European Particle Accelerator Conference (EPAC08)}}},\ \bibinfo {editor} {edited by\ \bibinfo {editor} {\bibfnamefont {P.}~\bibnamefont {Pierini}}}\ (\bibinfo {address} {Genoa, Italy},\ \bibinfo {year} {2008})\ p.\ \bibinfo {pages} {MOPP128}\BibitemShut {NoStop}%
\bibitem [{\citenamefont {Argyropoulos}\ \emph {et~al.}(2018)\citenamefont {Argyropoulos}, \citenamefont {Catalan-Lasheras}, \citenamefont {Grudiev}, \citenamefont {Mcmonagle}, \citenamefont {Rodriguez-Castro}, \citenamefont {Syrachev}, \citenamefont {Wegner}, \citenamefont {Woolley}, \citenamefont {Wuensch}, \citenamefont {Zha}, \citenamefont {Dolgashev}, \citenamefont {Bowden}, \citenamefont {Haase}, \citenamefont {Lucas}, \citenamefont {Volpi}, \citenamefont {Esperante-Pereira},\ and\ \citenamefont {Rajam\"aki}}]{PhysRevAccelBeams.21.061001}%
  \BibitemOpen
  \bibfield  {author} {\bibinfo {author} {\bibfnamefont {T.}~\bibnamefont {Argyropoulos}}, \bibinfo {author} {\bibfnamefont {N.}~\bibnamefont {Catalan-Lasheras}}, \bibinfo {author} {\bibfnamefont {A.}~\bibnamefont {Grudiev}}, \bibinfo {author} {\bibfnamefont {G.}~\bibnamefont {Mcmonagle}}, \bibinfo {author} {\bibfnamefont {E.}~\bibnamefont {Rodriguez-Castro}}, \bibinfo {author} {\bibfnamefont {I.}~\bibnamefont {Syrachev}}, \bibinfo {author} {\bibfnamefont {R.}~\bibnamefont {Wegner}}, \bibinfo {author} {\bibfnamefont {B.}~\bibnamefont {Woolley}}, \bibinfo {author} {\bibfnamefont {W.}~\bibnamefont {Wuensch}}, \bibinfo {author} {\bibfnamefont {H.}~\bibnamefont {Zha}}, \bibinfo {author} {\bibfnamefont {V.}~\bibnamefont {Dolgashev}}, \bibinfo {author} {\bibfnamefont {G.}~\bibnamefont {Bowden}}, \bibinfo {author} {\bibfnamefont {A.}~\bibnamefont {Haase}}, \bibinfo {author} {\bibfnamefont {T.~G.}\ \bibnamefont {Lucas}}, \bibinfo {author} {\bibfnamefont {M.}~\bibnamefont {Volpi}}, \bibinfo {author} {\bibfnamefont
  {D.}~\bibnamefont {Esperante-Pereira}},\ and\ \bibinfo {author} {\bibfnamefont {R.}~\bibnamefont {Rajam\"aki}},\ }\bibfield  {title} {\bibinfo {title} {Design, fabrication, and high-gradient testing of an $x$-band, traveling-wave accelerating structure milled from copper halves},\ }\href {https://doi.org/10.1103/PhysRevAccelBeams.21.061001} {\bibfield  {journal} {\bibinfo  {journal} {Phys. Rev. Accel. Beams}\ }\textbf {\bibinfo {volume} {21}},\ \bibinfo {pages} {061001} (\bibinfo {year} {2018})}\BibitemShut {NoStop}%
\bibitem [{\citenamefont {{Dassault Systèmes}}(2023)}]{CSTStudioSuite2023}%
  \BibitemOpen
  \bibfield  {author} {\bibinfo {author} {\bibnamefont {{Dassault Systèmes}}},\ }\href {https://www.3ds.com/products/simulia/cst-studio-suite} {\bibinfo {title} {{CST} studio suite}} (\bibinfo {year} {2023})\BibitemShut {NoStop}%
\bibitem [{\citenamefont {Seibel}(2017)}]{Seibel2017}%
  \BibitemOpen
  \bibfield  {author} {\bibinfo {author} {\bibfnamefont {A.}~\bibnamefont {Seibel}},\ }\emph {\bibinfo {title} {Untersuchungen zu einer neuen Alvarez-Struktur f{\"u}r den GSI Post-Stripper}},\ \href@noop {} {\bibinfo {type} {doctoralthesis}},\ \bibinfo  {school} {Johann Wolfgang Goethe-Universität Frankfurt am Main} (\bibinfo {year} {2017})\BibitemShut {NoStop}%
\bibitem [{\citenamefont {Du}\ \emph {et~al.}(2017)\citenamefont {Du}, \citenamefont {Groening}, \citenamefont {Mickat}, \citenamefont {Seibel},\ and\ \citenamefont {Kester}}]{PhysRevAccelBeams.20.032001}%
  \BibitemOpen
  \bibfield  {author} {\bibinfo {author} {\bibfnamefont {X.}~\bibnamefont {Du}}, \bibinfo {author} {\bibfnamefont {L.}~\bibnamefont {Groening}}, \bibinfo {author} {\bibfnamefont {S.}~\bibnamefont {Mickat}}, \bibinfo {author} {\bibfnamefont {A.}~\bibnamefont {Seibel}},\ and\ \bibinfo {author} {\bibfnamefont {O.~K.}\ \bibnamefont {Kester}},\ }\bibfield  {title} {\bibinfo {title} {Field stabilization of alvarez-type cavities},\ }\href {https://doi.org/10.1103/PhysRevAccelBeams.20.032001} {\bibfield  {journal} {\bibinfo  {journal} {Phys. Rev. Accel. Beams}\ }\textbf {\bibinfo {volume} {20}},\ \bibinfo {pages} {032001} (\bibinfo {year} {2017})}\BibitemShut {NoStop}%
\bibitem [{\citenamefont {Du}\ and\ \citenamefont {Groening}(2018)}]{PhysRevAccelBeams.21.084601}%
  \BibitemOpen
  \bibfield  {author} {\bibinfo {author} {\bibfnamefont {X.}~\bibnamefont {Du}}\ and\ \bibinfo {author} {\bibfnamefont {L.}~\bibnamefont {Groening}},\ }\bibfield  {title} {\bibinfo {title} {Compression and noise reduction of field maps},\ }\href {https://doi.org/10.1103/PhysRevAccelBeams.21.084601} {\bibfield  {journal} {\bibinfo  {journal} {Phys. Rev. Accel. Beams}\ }\textbf {\bibinfo {volume} {21}},\ \bibinfo {pages} {084601} (\bibinfo {year} {2018})}\BibitemShut {NoStop}%
\bibitem [{\citenamefont {Bethe}\ and\ \citenamefont {Schwinger}(1943)}]{BetheSchwinger1943PerturbationTheory}%
  \BibitemOpen
  \bibfield  {author} {\bibinfo {author} {\bibfnamefont {H.~A.}\ \bibnamefont {Bethe}}\ and\ \bibinfo {author} {\bibfnamefont {J.}~\bibnamefont {Schwinger}},\ }\href@noop {} {\emph {\bibinfo {title} {Perturbation theory for cavities}}},\ \bibinfo {type} {N.D.R.C. Report}\ \bibinfo {number} {D1-117}\ (\bibinfo  {institution} {Cornell University},\ \bibinfo {year} {1943})\BibitemShut {NoStop}%
\bibitem [{\citenamefont {Dolgashev}\ \emph {et~al.}(2021)\citenamefont {Dolgashev}, \citenamefont {Faillace}, \citenamefont {Spataro}, \citenamefont {Tantawi},\ and\ \citenamefont {Bonifazi}}]{PhysRevAccelBeams.24.081002}%
  \BibitemOpen
  \bibfield  {author} {\bibinfo {author} {\bibfnamefont {V.~A.}\ \bibnamefont {Dolgashev}}, \bibinfo {author} {\bibfnamefont {L.}~\bibnamefont {Faillace}}, \bibinfo {author} {\bibfnamefont {B.}~\bibnamefont {Spataro}}, \bibinfo {author} {\bibfnamefont {S.}~\bibnamefont {Tantawi}},\ and\ \bibinfo {author} {\bibfnamefont {R.}~\bibnamefont {Bonifazi}},\ }\bibfield  {title} {\bibinfo {title} {High-gradient rf tests of welded $x$-band accelerating cavities},\ }\href {https://doi.org/10.1103/PhysRevAccelBeams.24.081002} {\bibfield  {journal} {\bibinfo  {journal} {Phys. Rev. Accel. Beams}\ }\textbf {\bibinfo {volume} {24}},\ \bibinfo {pages} {081002} (\bibinfo {year} {2021})}\BibitemShut {NoStop}%
\bibitem [{\citenamefont {Steele}(1966)}]{1126168}%
  \BibitemOpen
  \bibfield  {author} {\bibinfo {author} {\bibfnamefont {C.}~\bibnamefont {Steele}},\ }\bibfield  {title} {\bibinfo {title} {A nonresonant perturbation theory},\ }\href {https://doi.org/10.1109/TMTT.1966.1126168} {\bibfield  {journal} {\bibinfo  {journal} {IEEE Transactions on Microwave Theory and Techniques}\ }\textbf {\bibinfo {volume} {14}},\ \bibinfo {pages} {70} (\bibinfo {year} {1966})}\BibitemShut {NoStop}%
\bibitem [{\citenamefont {Wegner}\ \emph {et~al.}(2014)\citenamefont {Wegner}, \citenamefont {Wuensch}, \citenamefont {Burt},\ and\ \citenamefont {Woolley}}]{Wegner2014BeadPullMethod}%
  \BibitemOpen
  \bibfield  {author} {\bibinfo {author} {\bibfnamefont {R.}~\bibnamefont {Wegner}}, \bibinfo {author} {\bibfnamefont {W.}~\bibnamefont {Wuensch}}, \bibinfo {author} {\bibfnamefont {G.}~\bibnamefont {Burt}},\ and\ \bibinfo {author} {\bibfnamefont {B.}~\bibnamefont {Woolley}},\ }\bibfield  {title} {\bibinfo {title} {Bead-pull measurement method and tuning of a prototype clic crab cavity},\ }in\ \href {https://accelconf.web.cern.ch/LINAC2014/papers/MOPP035.pdf} {\emph {\bibinfo {booktitle} {Proceedings of LINAC2014}}},\ \bibinfo {editor} {edited by\ \bibinfo {editor} {\bibfnamefont {C.}~\bibnamefont {Carli}}}\ (\bibinfo  {publisher} {JACoW},\ \bibinfo {address} {Geneva, Switzerland},\ \bibinfo {year} {2014})\ p.\ \bibinfo {pages} {MOPP035}\BibitemShut {NoStop}%
\bibitem [{\citenamefont {Heilmann}\ \emph {et~al.}(2018)\citenamefont {Heilmann}, \citenamefont {Du}, \citenamefont {Groening}, \citenamefont {Kaiser}, \citenamefont {Mickat}, \citenamefont {Seibel},\ and\ \citenamefont {Vossberg}}]{Heilmann2018ScaledAlvarez}%
  \BibitemOpen
  \bibfield  {author} {\bibinfo {author} {\bibfnamefont {M.}~\bibnamefont {Heilmann}}, \bibinfo {author} {\bibfnamefont {X.}~\bibnamefont {Du}}, \bibinfo {author} {\bibfnamefont {L.}~\bibnamefont {Groening}}, \bibinfo {author} {\bibfnamefont {M.}~\bibnamefont {Kaiser}}, \bibinfo {author} {\bibfnamefont {S.}~\bibnamefont {Mickat}}, \bibinfo {author} {\bibfnamefont {A.}~\bibnamefont {Seibel}},\ and\ \bibinfo {author} {\bibfnamefont {M.}~\bibnamefont {Vossberg}},\ }\bibfield  {title} {\bibinfo {title} {Scaled alvarez-cavity model investigations for the unilac upgrade},\ }in\ \href {https://doi.org/10.18429/JACoW-IPAC2018-TUPAF079} {\emph {\bibinfo {booktitle} {9th International Particle Accelerator Conference (IPAC2018)}}},\ \bibinfo {editor} {edited by\ \bibinfo {editor} {\bibfnamefont {S.}~\bibnamefont {Koscielniak}}}\ (\bibinfo {organization} {JACoW Publishing},\ \bibinfo {address} {Vancouver, BC, Canada},\ \bibinfo {year} {2018})\ p.\ \bibinfo {pages} {TUPAF079}\BibitemShut {NoStop}%
\bibitem [{\citenamefont {Podlech}\ \emph {et~al.}(2007)\citenamefont {Podlech}, \citenamefont {Ratzinger}, \citenamefont {Klein}, \citenamefont {Commenda}, \citenamefont {Liebermann},\ and\ \citenamefont {Sauer}}]{PhysRevSTAB.10.080101}%
  \BibitemOpen
  \bibfield  {author} {\bibinfo {author} {\bibfnamefont {H.}~\bibnamefont {Podlech}}, \bibinfo {author} {\bibfnamefont {U.}~\bibnamefont {Ratzinger}}, \bibinfo {author} {\bibfnamefont {H.}~\bibnamefont {Klein}}, \bibinfo {author} {\bibfnamefont {C.}~\bibnamefont {Commenda}}, \bibinfo {author} {\bibfnamefont {H.}~\bibnamefont {Liebermann}},\ and\ \bibinfo {author} {\bibfnamefont {A.}~\bibnamefont {Sauer}},\ }\bibfield  {title} {\bibinfo {title} {Superconducting ch structure},\ }\href {https://doi.org/10.1103/PhysRevSTAB.10.080101} {\bibfield  {journal} {\bibinfo  {journal} {Phys. Rev. ST Accel. Beams}\ }\textbf {\bibinfo {volume} {10}},\ \bibinfo {pages} {080101} (\bibinfo {year} {2007})}\BibitemShut {NoStop}%
\end{thebibliography}%

\end{document}